# Ease on Down the Code: Complex Collaborative Qualitative Coding Simplified with 'Code Wizard'

ABBAS GANJI, University of Washington, USA
MANIA ORAND, Institute for Research in Fundamental Sciences (IPM) & University of Washington, USA
DAVID W. MCDONALD, University of Washington, USA

This paper describes the design and development of a preliminary qualitative coding tool as well as a method to improve the process of achieving inter-coder reliability (ICR) in small teams. Software applications that support qualitative coding do not sufficiently assist collaboration among coders and overlook some fundamental issues related to ICR. We propose a new dimension of collaborative coding called "coders' certainty" and demonstrate its ability to illustrate valuable code disagreements that are missing from existing approaches. Through a case study, we describe the utility of our tool, Code Wizard, and how it helped a group of researchers effectively collaborate to code naturalistic observation data. We report the valuable lessons we learned from the development of our tool and method: (1) identifying coders' certainty constitutes an important part of determining the quality of data analysis and facilitates identifying overlapping and ambiguous codes, (2) making the details of coding process visible helps streamline the coding process and leads to a sense of ownership of the research results, and (3) there is valuable information hidden in coding disagreements that can be leveraged for improving the process of data analysis.

CCS Concepts: • **Human-centered computing → Collaborative and social computing theory, concepts and paradigms**

**KEYWORDS**

Qualitative coding, collaborative coding, qualitative coding software application, qualitative code disagreements

**132**

## 1 INTRODUCTION

Collaborative qualitative coding is an iterative and often lengthy process mainly consisting of reading, interpreting, classifying, and analyzing qualitative data such as interview transcripts and field notes. The length of the coding process is correlated primarily with the volume of such data and the number of individual collaborating coders. There are various approaches to conduct collaborative coding inspired by different philosophical paradigms such as positivism and interpretivism.

Author's addresses: A. Ganji, Department of Civil and Environment Engineering, University of Washington, Seattle, USA; M. Orand, School of Computer Science, Institute for Research in Fundamental Sciences (IPM), and Department of Human-Centered Design & Engineering, University of Washington, Seattle, USA; D. McDonald, Department of Human-Centered Design & Engineering, University of Washington, Seattle, USA.

In this paper, we focus on the positivist perspective, which entails individual coders coding the same data independently and separately, and then comparing the results to assess inter-coder reliability (ICR).

For large volumes of data, it is critical to have multiple coders collaborate to establish trustworthiness of the data analysis and decrease coding errors [1-6]. Krippendorff states that a minimum of three coders is required for any collaborative qualitative coding reliability and to achieve "confidence in the data beyond the measured reliability" [7]. However, as we discuss in detail in Section 2, collaboration among multiple coders is often challenging, because (1) coordinating and organizing multiple coders and several rounds of coding is complex [8], (2) achieving an acceptable ICR can be complicated [7-9] because ICR is impacted by factors such as number of coders, coders' skills, data size, as well as whether coders have the same understanding and interpretation of data, categories, or instructions [10], (3) when ICR is low, finding the reasons and addressing the problems is often arduous, because it requires backtracking from aggregated codes to find the individual codes and then discussing the related quotes with the coders, and (4) coders can agree with each other but still be wrong [1 and 12].

Software tools such as NVivo, Atlas.ti, Dedoose, HyperResearch, and SaturateApp are available to support qualitative coding. However, as we discuss in Section 2, despite their important features and a clear need for these tools, they are not broadly adopted for collaborative qualitative coding, especially in smaller groups and academic settings [13].

Many researchers often rely on spreadsheets for collaborative qualitative coding. As effective as they may be, spreadsheets are not designed for this purpose and offer few features required in qualitative coding. For example, Google Spreadsheets facilitate multi-user collaboration in real-time, but require careful management of access controls or the coded data by each individual coder would be visible to all other coders, which violates some principles of qualitative coding. Alternatively, offline spreadsheets provide an unaided coding experience and are not specifically helpful in aggregating and processing the coded data.

Qualitative researchers require different dimensions of support. Unaided collaborative coding is inefficient and time consuming, while commercially available coding tools are expensive, require training, and are mostly useful for projects where the coding scheme is known upfront. We feel there is a need for tools that would support middle-level coding needs. Thus, our motivation was to develop a lightweight tool to help middle-level collaborative qualitative coding projects with very low adoption overhead.

We would like to emphasize that our goal is not to replace the existing commercial coding tools. These sophisticated tools support a wide range of data types including text, audio, images, and video, and offer expansive functionality. Our tool, in its current state, is designed to support the collaborative aspects of coding textual data within small teams.

We describe the development process of our preliminary collaborative coding tool, called Code Wizard, focusing on facilitating collaboration through visualization. We programmed and developed Code Wizard's prototype on Microsoft Excel (Excel) over the course of 30 weeks. Unlike most qualitative coding tools, Code Wizard requires minimal training because it appears to its users as a basic spreadsheet, a tool already quite familiar to many qualitative researchers.

Code Wizard supports a team's collaboration in three main ways: (1) it facilitates the discovery of ambiguous and overlapping codes, (2) it enhances team discussions and the reconciliation process by generating real-time results and allowing immediate access to note excerpts, codebooks, and other details, and (3) it accelerates the process of coming to agreement through color coding and visualizing discrepancies.

In addition, we introduce a new dimension in collaborative coding, "coders' certainty," and discuss how it improved the collaborative process of achieving coders' reliability in a small team. We describe our method of leveraging correlated disagreements to support achieving higher agreement threshold (i.e., higher ICR). We also demonstrate Code Wizard's utility by discussing how it helped a group of researchers collaborate on coding observation field notes.

It is important to point out that qualitative coding includes various components, but in this paper, we focus exclusively on the issues related to collaborative coding from positivist perspective, and not other aspects such as segmentation/unitization, and developing categories and themes.

The following research questions guided our research:

RQ1: What aspects of collaborative qualitative coding can facilitate the challenges of collaboration?

RQ2: How can we use these aspects to design a tool to facilitate the collaborative coding process?

We first review the literature of qualitative data analysis with special attention to collaborative coding. We briefly review a few popular qualitative coding software applications. Then, we describe our methodology and the development of Code Wizard and demonstrate its application through a case study. We conclude by discussing our findings, recognizing areas for improvements, and identifying directions for future research.

## 2 BACKGROUND AND RELATED WORK

### 2.1 Collaborative Coding

Collaborative coding is key to providing sound interpretation of qualitative data and is vital to creating valid results. Burla et al. (2008) indicate that for large volumes of data, it is best to have more than one person code for efficiency as it helps increase comprehension, support inter-subjectivity, and provide sound interpretation of the data [3].

Collaborative coding with multiple coders benefits from a diversity of knowledge and skills that can result in more clarity in codes, the main cause of disagreement. The disagreements are often addressed through clarification discussions that may lead to refining coding frame [12 & 14].

Collaborative coding is a response to the challenges of subjectivity of qualitative data analysis. Barbour (2001) indicates that "the greatest potential of collaborative coding lies in its capacity to furnish alternative interpretations and thereby to act as the 'devil's advocate' implied in many of the checklists

> **Research Project**
>
> In June 2016, the largest regional disaster response exercise conducted in the Pacific Northwest took place. Cascadia Rising 2016 was a "four-day, large scale exercise to test response and recovery capabilities in the wake of a 9.0 magnitude CSZ earthquake and tsunami." The exercise involved local, state, tribal, federal partners, nongovernmental organizations, the private sector, and military commands.
>
> Over 20,000 emergency managers participated in the exercise in Washington, Idaho, and Oregon [25-27].
>
> Our team saw this exercise as an opportunity to study a complex, dynamic, and non-routine event and investigate the obstacles of information sharing and coordination during a disaster event. Our team conducted fieldwork observations at Emergency Operations and Coordination Centers, and the Federal Emergency Management Agency and major military centers for 4 days and collected 8 to 10 hours of field notes per day. Field observations were conducted at state and county locations.
>
> **Sidebar 1**

in alerting researchers to all potentially competing explanations." She emphasizes the value of collaborative coding because it leads researchers to confront different interpretations of data or codebook in research team meetings, which improves the quality of the qualitative content analysis [14].

### 2.2 Inter-coder Reliability

Krippendorff (2004, 2011) defines reliability as "the degree to which members of a designated community agree on the readings, interpretations, responses to, or uses of given texts or data." While reliability has different components, reproducibility (interchangeably used as inter-coder reliability or interrater reliability) is the minimum requirement to establish reliability in qualitative data analysis. Reproducibility is defined as "the degree to which a process can be replicated by different analysts working under varying conditions, at different locations, or using different but functionally equivalent measuring instruments" [1 and 7].

According to Lombard et al. (2002), ICR is widely used to identify "the extent to which independent coders evaluate a characteristic of a message or artifact and reach the same conclusion" [11]. Barbour (2001) highlights the importance of taking a qualitative approach to identify reasons behind coding disagreements and not solely rely on quantitatively measuring ICR. The authors support the "merit of coding discussions" to improve qualitative analysis [14].

While many people use the terms reliability and agreement interchangeably, Krippendorff (2004) notes, "agreement is what we measure; reliability is what we wish to infer from it. In content analysis, reproducibility is arguably the most important interpretation of reliability" [7]. Lombard et al. (2002) also state "[i]t is widely acknowledged that intercoder reliability is a critical component of content analysis and (although it does not ensure validity) when it is not established, the data and interpretations of the data can never be considered valid." ICR can be inferred by measuring agreement level among coders in collaborative analysis [11].

Artstein and Poesio (2008) compared different methods of measuring agreement among coders in collaborative coding such as direct percentage of agreement, Scott's pi, Cohen's kappa, Krippendorff's alpha, and Fleiss' kappa. They do not recommend direct percentage of agreement because it does not consider the chance agreement in its formula. The direct percentage method is also limited to two coders. Except for Krippendorff's alpha and Fleiss' kappa, other methods fail in measuring agreement for more than two coders. Fleiss' kappa is

---

We used Slack [28] to capture observations because it (a) time-stamps the notes, (b) facilitates communication between the observers, (c) has a built-in search feature, and (d) allows researchers at multiple sites to follow and track data collection in real-time and help clarify data when necessary.

The coding team consisted of 5 graduate and undergraduate students from the Departments of Civil Engineering, Cognitive Psychology, Electrical Engineering, and Human-Centered Design & Engineering. An experienced qualitative researcher led the coding process.

Our coding team faced 4 main challenges: (1) the coders were not familiar with disaster recovery; (2) team members changed quarterly, and onboarding new coders every 10-12 weeks was not realistic. Tools such as Dedoose and NVivo require training [19]; (3) the coders' diverse backgrounds and varying experiences in qualitative coding necessitated extra work to reach a common language; and (4) The team found no software that they all were proficient at or could be mastered quickly.

**Sidebar 2**

a generalized form of Scott's pi for multiple coders and also much simpler than Krippendorff's alpha approach [15].

Although there is no standard threshold for determining acceptable level of reliability [11], Fleiss' kappa of higher than 0.8 is considered an "almost perfect" level of agreement among coders [11, 16].

### 2.3 Challenges of Achieving High ICR

Achieving an acceptable ICR is impacted by various factors. According to Hruschka et al. (2004), a large sample size may increase the complexity of coding and decrease ICR. In addition, "variation in the content, length of response, or number of codes per question may affect the speed at which a team achieves an acceptable level of inter-coder reliability." Finally, "variation in the clarity of the codebook and individual code definitions" are other factors that can influence the reliability process [9].

The use of multiple coders may introduce new challenges. According to Berends and Johnston (2005), in a team of multiple coders, different skill levels, resources, and time demands can slow down the collaborative coding process. However, there is a tradeoff to consider, because inclusion of multiple perspectives in researcher backgrounds plays an important role in discussing the disagreements and refining the coding system. Similarly, Barbour (2001) states that engaging multiple coders improves the clarity of the codebook and helps identify reasons behind coding disagreements [12 and 14].

Burla et al. (2008) categorize the sources for low ICR into two groups: (1) factors related to codes such as weaknesses of coding scheme, definitions, examples, and classification rules, (when codes were mutually exclusive), and (2) factors related to coders' skill levels such as training issues that lead to insufficient understanding or inappropriate application of the coding scheme by one coder [3].

Collaborating remotely with the research team is another reason for low ICR. MacPhail et al. (2016) report that their team had to conduct 10 rounds of collaborative coding over the course of one year to achieve high Cohen's kappa [8]. They state that their main challenge was coders' different understanding of how to use and apply the codes because the coders were located in different countries. Coders needed to use phone and emails to communicate with the rest of the team, which lengthened the coding process. MacPhail et al. also state that "there are limited practical resources available for researchers engaging in a group coding process and interested in ensuring adequate Inter-coder Reliability (ICR); the amount

> Our research team was required to complete an online disaster recovery and management course offered by the government when they joined the project, which helped with the first challenge.
>
> However, quarterly changes to team composition made the coding process challenging because new team members needed time to understand the codes and each other's perspectives, which was time-consuming.
>
> When recruiting new project team members each quarter, we did not want to recruit coders based on their knowledge of particular tools. We also did not want to compromise the data collection, data usage, and data analysis by tailoring the study needs around the capabilities of particular software tools.
>
> As Woods et al. [19] indicate, "software can dominate the researchers' understanding of their practices" especially if the qualitative data analysis software tools are used "without a critical and reflexive awareness" of their influences on qualitative research practices.
>
> **Sidebar 3**

of agreement between two or more coders for the codes applied to qualitative text." MacPhail et al. report that they had to use Coding Analysis Toolkit (CAT) [17] in addition to Atlas.ti Version 7.0 because Atlas.ti did not support ICR statistics [18].

### 2.4 Software Tools and Visualization Techniques for Qualitative Coding

There are various software applications to support collaborative qualitative coding. Peter Nielsen (2012) evaluated and compared the features of four prominent tools including, NVivo, Atlas.ti, Dedoose, and HyperResearch. Nielsen demonstrated that NVivo, Atlas.ti, and HyperResearch are more useful for individual coding but "are insufficient to support collaborative coding." On the other hand, Dedoose is "fully and transparently collaborative," but it lacks certain features to support data analysis and "does not easily relate codes to each other in a structure (hierarchical or network)" [13].

Woods et al. [19] investigated 763 empirical articles to explore how researchers use Atlas.ti and NVivo and found that although these tools can support multiple phases of the research process, the majority of researchers use them for data management and analysis, "with fewer using it for data collection/creation or to visually display their methods and findings."

Some researchers use visuals and colors to help the coding process. For example, Blascheck et al. use a visual analytics approach to integrate various data formats, such as transcripts and videos [22]. The authors create a tabular representation of the data using colors, panels, and Sidebars to provide an overview of all user activities. Quirkos [23] uses colors to categorize and cluster codes and are more focused on individual coding. Aeonium [24] is a coding interface that uses machine learning and visual analytic techniques to facilitate identifying ambiguous data. Aeonium helps direct coders' attention to the ambiguous data that warrants analysis and is more focused on the collaboration aspect of coding.

## 3 METHODOLOGY

In this section, we describe the iterative process of developing Code Wizard and illustrate its application to our research project, which is described in the Sidebars. The Sidebars narrate a case study in which we realized the need to improve our collaborative coding methodology and develop a new tool to address our challenges. Sidebar 1 describes the context of our research project. Sidebar 2 gives insight into our data

---

The team made initial attempts to use Dedoose [21] based on the recommendation of our lead researcher. However, we abandoned Dedoose after a few weeks because (1) the team members felt Dedoose was difficult and time-consuming to learn, (2) it was slow to work with, especially for the code disagreement discussion meetings, and (3) it did not offer the data analysis functionalities the team required, such as calculating Fleiss' kappa or other ICR measures for more than two coders.

The team decided not to switch to a new tool such as NVivo or Atlas.ti, because in addition to our unsatisfactory experience with Dedoose, we did not have time to obtain and learn a new tool.

We realized Microsoft Excel was well known to all members of the team and, as a baseline tool, could be easily adopted. However, Excel is not a coding tool and initially the team mostly used it to organize the codes. Experiencing the challenges of an unaided coding process within our small team, we felt the need to add more features to Excel to facilitate our collaboration process ever so slightly.

**Sidebar 4**

collection procedures, coding team characteristics, and the main challenges that our coding team faced. Sidebar 3 discusses the challenges caused by the structure of our research team.

### 3.1 Research Method

The design and development of Code Wizard had humble beginnings. As we describe in Sidebar 4, our team decided to use Excel to code the textual field notes. We soon realized our coding efforts could benefit from incorporating a few additional features to our Excel sheets, for example to help sort the coded data for team discussions or to produce graphs. Initially, one team member led the design and development of a module by adding features to Excel spreadsheets to address the immediate needs of the coders. Through iterative cycles of use, coders' feedback, and redesign, the tool gradually evolved into a piece of software (See Code 1).

While this design and development process may not seem systematic, our approach has some of the characteristics of practice-based research [29]. We were looking for a way to address our constantly changing environment, had identified some problems (discussed in Section 3.3), saw the need for a tool to support collaborative coding, and started to build such a tool. We deployed the tool in our small coding team and observed how it impacted the team and each individual coder. We then iteratively refined and tested the tool based on the feedback provided by team members or by testing it live in our team discussions.

During the design and development of Code Wizard, we collected three types of data:

(1) Data about the tool: we have every version of Code Wizard. This helped us track the changes, show how we modified the tool over time, and understand the reasons that warranted a software update.

(2) Data on the team's coding activities: we collected all versions of the coding activities. These data were important because: (a) the coded data over time provided a way to trace how the coding evolved and how the accuracy or inaccuracy of codes changed as a result of the relationship of the tool with the team, and (b) the applied codes explicitly contributed to the actual disaster recovery research.

(3) Data about the team: team members provided verbal feedback and the developer was present in all of the coding and debriefing discussions as the tool was evolving. To triangulate the verbal feedback, we have access to the emails sent by the coding team to the developer advocating for the changes and modifications of the tool. In addition, some of the team members provided written feedback at the end of the project. For example, one of the coders wrote, "[...] the [Excel spread]sheets were better in helping us compare coder's codes temporally within one screen." Another coder wrote, "The sheets worked well with less overhead than Dedoose, which was the platform we originally tried using for the project."

Data Analysis: The primary analysis was to take verbal and written feedback from the team and interpret it such that the tool could resolve the raised issues by the team members.

### 3.2 What is Code Wizard?

As described in Sidebar 2, our team initially experienced four main challenges: (1) the domain of our study was disaster recovery and management, and not all of the coders were familiar with that domain; (2) the coding team consisted of 5 graduate and undergraduate students from various backgrounds. However, the team members changed quarterly, and onboarding new coders every 10-12 weeks was not realistic. This was important because tools such as Dedoose and NVivo are sophisticated and require training [19]; (3) the coders' diverse backgrounds and varying experiences in qualitative coding necessitated extra work to reach a common language, and (4) the team found no software that they all were proficient at or could be mastered quickly.

In addition to these 'team and software-related' challenges, our team identified three 'process-related' challenges, described in Section 3.3. We realized that our team required some software features that were either missing from the existing tools for collaborative coding or were too complex to learn for our project. Our team's fallback plan was to use Excel to organize the data and codes, and from there, we gradually programmed Excel based on the unfolding needs of the team.

Code Wizard is a visualization tool embedded in Excel.[1] It initially consisted of two Excel files: a pre-formatted spreadsheet for individual coders to use independently, and another spreadsheet that aggregates the individual coding spreadsheets and automatically sorts and compares coded data from individual coders, calculates ICR, and produces visualizations. The aggregated spreadsheet is primarily for the coding lead to use after individual coders finish coding. Below, we discuss each of these files in details.

**Code 1:** An example of a Visual Basic pseudo code that generates Figs. 4, 5, and 7

```
# Aggregating codes assigned to units of analysis by different coders (primary & secondary)
for coder in codersList:
  primaryCoded += primaryCodedUnitsBy(coder)
  secondaryCoded += secondaryCodedUnitsBy(coder)
# Calculating correlation between primary and secondary codes
for row in codebook:
  for column in codebook:
    i = 0
    nPri = 0
    nSec = 0
    for unit in primaryCoded:
      if unit is row:
        # Aggregates total uses of a code as primary
        nPri += 1
        if secondaryCoded [ i ] is column:
          # Aggregates total uses of a code as secondary for a given primary code
          nSec += 1
      i += 1
    codeCorrelation = nSec/nPri
```

The individual coding file includes three spreadsheets. The first is the main spreadsheet tab where the units of analysis are loaded and sorted in rows and the codes (categories) are organized in a dropdown list. Code Wizard designates one column for primary codes and one column for secondary codes. We will discuss the primary and secondary codes in Section 3.3. Coders can click on the dropdown lists in front of each unit of analysis to select the codes. There are hyperlinks in front of each unit of analysis that take the coders to the source of textual data if they need it. In addition, a "Help" hyperlink directs the coders to the codebook instructions when needed. The second spreadsheet tab includes the codebook and instructions.

---

[1] Code Wizard is available on www.codewizard.online

Code Wizard assigns a unique color to each code. Finally, the third spreadsheet tab includes the textual data as a reference (Fig. 1).

| Codes | Definition |
|---|---|
| Computational Mechanism Issue (CM) | Perhaps poorly designed computer-based form or system |
| Content Breakdown (CT) | Unclear or incomplete content; missing information |
| Coordination Breakdown (CD) | Individuals/groups knew the process, but process did not work, or they did not follow it |
| Disparate Systems (DS) | Systems do not work well together; may occur when two systems have two different standards |
| Paper Mechanism Issue (PM) | Paper form not working, perhaps due to poor design or missing information |
| Source Breakdown (SB) | Source not identified; whenever we think something went wrong regardless of paper or system |
| Unclear Process (UP) | People unsure how to do things due to unclear or undocumented process including the lack of training |

Fig 1. An example of a colored codebook

The aggregated spreadsheet file is primarily used by the coding lead. After the individual coders finish coding, they submit it to the lead. Then, the lead collects all the individual spreadsheets and copies the codes into the aggregated sheet. Code Wizard will then automatically organize and sort the codes, calculate the ICR, and generate the visualizations. Similar to the individual spreadsheets, the aggregated sheet includes a tab for the codebook, a tab for each individual coder's coded data, a tab for the source of data, and two tabs for visualizations. These features and tabs help the lead and the team during discussion sessions.

Fig. 3 illustrates the first tab in the aggregated sheet, and Figs. 4, 7, 12, and 13 are examples of the tabs with visualizations. As the team continued to use and refine the tool, Code Wizard evolved to a Microsoft Excel Module. Code 1 shows an example of a Visual Basic pseudo code that generates Figs. 4, 5, and 7 in the aggregated spreadsheet file.

### 3.3 Collaborative Coding Challenges and Solutions

In this section, we describe each challenge we encountered during our collaborative coding process, followed by our solution. Fig. 1 shows our codebook, which includes seven mutually exclusive codes. We will discuss our results in Section 4.

*3.3.1 First Challenge: Discovering Ambiguous Codes.* The initial discussions among coders and the results of sample coding revealed that coders were not fully certain about the codes they had assigned to each unit of analysis. Some of the codes had created more confusion than others. However, the sheer act of assigning one code to a unit of analysis conceals the uncertainties behind that code. We believe there is a difference between assigning a code with high confidence and assigning a code because no other code fits.

As noted previously in Section 2, current methods of measuring ICR reduce the root causes of disagreements to a simple measurement of ICR across all coders. Thus, we were determined to further explore the reasons for coding discrepancies and disagreements and find a way to discern the ambiguous codes and evaluate the degree to which those codes had caused confusion.

*Addressing the First Challenge: Coders' Certainty, A Dimension to Support Collaborative Coding.* In the first round of coding, we asked individual coders to code each unit of analysis twice. We thought this double coding would help evaluate the uncertainty of individual coders as well as the entire team. The first code should be what the coders deemed as the best fit to the unit of analysis, which we call the primary code. A secondary code would show whether some different codes could potentially be assigned to that same unit of analysis. If coders were absolutely certain about their primary code, then we asked them to select the same code for the secondary code. Otherwise, coders assigned a different secondary code that would be a fit.

We prototyped this coding technique in a spreadsheet (Fig. 2) with the following main features: (1) we designated two separate columns for primary and secondary codes next to each other, and loaded all the codes to these columns and added a dropdown list to simplify the code selection process; (2) each code was colored differently to expedite locating the different primary and secondary codes; (3) for each unit of analysis, there was a direct link to the source data for coders' reference; (4) a "Help" button was included that linked to the codebook including the description of the codes, examples, instructions, and coding rules.

These features streamlined and accelerated the double coding process by allowing each coder to see all excerpts, access category definitions with one click, and select the codes from the drop down, pre-populated lists.

| | Time/Date | Unit of Analysis | Help? Codebook | Link to Source | Categories | |
|---|---|---|---|---|---|---|
| | | | | | Primary | Secondary |
| 1 | 10:08 6/7/16 | It is not clear if the information is from a citizen or someone else. | | go to | Source Breakdown | Source Breakdown |
| 2 | 10:08 6/7/16 | The agency that the information is coming from is not on the form. | | go to | Content Breakdown | Coordination breakdown |
| 3 | 10:08 6/7/16 | I ask about this and the player shows me other forms that similarly require a lot of guessing because the way they are filled out is vague. | | go to | Content Breakdown | Paper Mechanism Issue |
| 4 | 10:20 6/7/16 | One of the planning team goes to track down who handed off two confusing paper information forms. Guesses are made based on pen color and handwriting as to who to talk to. | | go to | Source Breakdown | Content Breakdown |

Fig. 2. A screenshot of the individual coding spreadsheet

When individual coders finished coding, our prototype aggregated all the codes and displayed them in one place, so the team could easily see the disagreements using visual cues (Fig. 3). For example, as can be seen in the first row in Fig. 3, coders have very high certainty about Source Breakdown (blue cells), with only one coder selecting a different secondary code for it. However, most coders were not certain about Content Breakdown (yellow cells) and selected a different secondary code for it.

We programmed this aggregated spreadsheet to calculate ICR using Fleiss' kappa statistical index. We preferred Fleiss' kappa method because (1) it is suitable when the number of coders exceeds two, (2) it is simpler and easier to understand compared to alternative methods such as Krippendorff's alpha, (3) it is a stronger method because, unlike many other methods that use direct percentage agreement, it considers the possibility of random agreement in its formula. An acceptable kappa should be equal or greater than 0.8 under Fleiss' method [15, 16].

| | Time | Unit of Analysis | Link to Source | John Primary | John Secondary | Ava Primary | Ava Secondary | Suzie Primary | Suzie Secondary | Robert Primary | Robert Secondary |
|---|---|---|---|---|---|---|---|---|---|---|---|
| 1 | 10:08 6/7/16 | It is not clear if the information is from a citizen or someone else. | go to | SB | CT | SB | SB | SB | SB | SB | SB |
| 2 | 10:08 6/7/16 | The agency that the information is coming from is not on the form. | go to | CT | CD | SB | SB | SB | SB | CT | SB |
| 3 | 10:08 6/7/16 | A player shows me other written forms that similarly require a lot of guessing because the way they are filled out is vague. | go to | CT | CD | CT | PM | CT | CT | CT | CT |
| 4 | 10:20 6/7/16 | A planning team member tracks down who handed off two confusing paper information forms. Guesses are made based on pen color and handwriting as to who to talk to. | go to | CT | CD | CT | SB | PM | CT | SB | SB |
| 5 | 10:20 6/7/16 | Another confusing message doesn't have a log number from the message center, so it was not routed through the center. So, the planning players are unsure who to go back and get information from. | go to | CT | CD | CT | SB | PM | CT | SB | SB |

(Help? Codebook)

Fig. 3. A screenshot of an aggregated spreadsheet. ICR=0.52

We only used primary codes to calculate ICR because Fleiss' kappa (and most other methods) is based on one code. Our ICR at the first round of coding was 0.52, which is below the acceptable kappa of 0.8. Thus, it was clear our team needed to resolve the disagreements.

After aggregating all the codes and calculating ICR, Code Wizard measured certainties by generating a graph illustrating the coders' certainty per code (Fig. 4). Code Wizard searched for all the individual units whose primary codes were similar, and then compared the primary with the secondary codes for each unit. For example, Code Wizard looked at all the units with Coordination Breakdown as the primary code and then checked whether the primary and secondary codes were similar. As can be seen in Fig. 4, the team had low certainty on Content Breakdown (56 percent), meaning that only 56 percent of the times where Content Breakdown was assigned as the primary code did the coders also assign it as the secondary code. On the other hand, the team was absolutely certain on using Coordination Breakdown, which means 100 percent of the times this code was selected as the primary code, it was also selected as the secondary code.

Fig. 4 was particularly useful because it highlighted the codes that needed further discussion, and it guided the team's discussions on the low certainty codes.

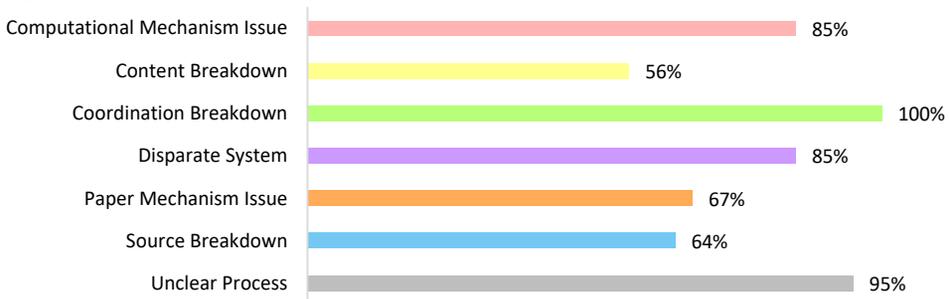

Fig. 4. Certainty of all coders per code

In addition, our prototype automatically generated the certainty graph for each individual coder (Fig. 5). For example, coder 2 was highly certain about Coordination Breakdown (Green bar) and Paper Mechanism Issue (Orange bar), but had low certainty on Content Breakdown and Disparate Systems. As Fig. 5 illustrates, coder 2 was absolutely certain about Paper Mechanism Issue (Orange bar), which is much higher than the team's certainty rate of 67%, as shown in Fig. 4. This information was helpful in discovering ambiguous codes.

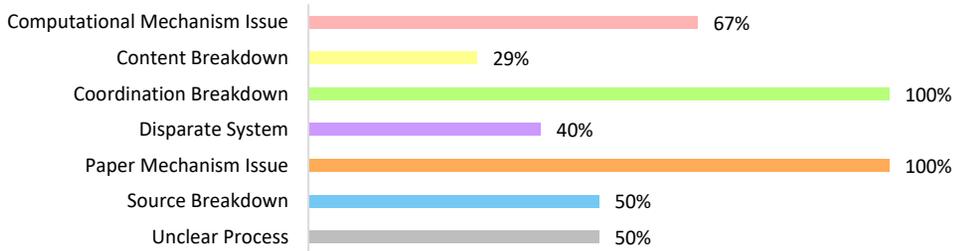

Fig. 5. Certainty of an individual coder per code – Coder 2

This graph helped the team in two ways: (1) individual coders could see where they struggled the most and determine how they needesd help with specific codes, and (2) individual coders could quickly see how their uncertainties compared to others and the team as a whole, and decide where to focus efforts to reconcile discrepancies.

Figs. 1 through 5 helped our team quickly and efficiently discover their uncertainties, disagreements, and the ambiguous codes. However, we needed more details to support the team and collaboration process more effectively.

*3.3.2 Second Challenge: Discovering the Overlapping Codes.* Although Figs. 1 through 5 revealed code ambiguities and showed overlap between some codes, they do not show which codes overlap and the degree to which these codes overlap. These details are important because in collaborative coding we aim for creating mutually exclusive codes to achieve higher agreement.

This problem is explained with an example in Fig. 6. For Unit 1, primary code A is assigned four times, and similarly, code B is assigned four times for Unit 2. From the perspective of existing coding approaches, Units 1 and 2 have the same level of agreement because four out of five coders assigned the same primary codes to these units of analysis.

However, there is more valuable information in Fig. 6. For Unit 1, the majority of coders assigned code C as the other possible code (Unit 1), but for Unit 2 secondary codes varied, such that codes A, B, C, and D were all used. For Unit 1, we can infer that there might be a strong overlap between codes A and C, especially if this pattern repeats in the coding results. However, code B might be too broad or defined poorly. We wanted to further explore these patterns and see how they could help our team reconcile the disagreements.

In existing collaborative coding methods, coders meet at this point to discuss the disagreements and attempt to find the ambiguous codes. They go over the coded data line by line to find the disagreements and discuss the rationale for their code assignments. Then, they decide whether to revise the code definitions, or take other actions such as adding, removing, or merging the codes. As noted earlier, without Code Wizard this step can take several months and many iterations of collaborative coding especially for large datasets with multiple coders [8, 9, and 17].

|  | Coder 1 | | Coder 2 | | Coder 3 | | Coder 4 | | Coder 5 | |
|---|---|---|---|---|---|---|---|---|---|---|
|  | Primary | Secondary | Primary | Secondary | Primary | Secondary | Primary | Secondary | Primary | Secondary |
| Unit 1 | **A** | C | C | A | **A** | C | **A** | C | **A** | C |
| Unit 2 | **B** | D | **B** | B | **C** | B | **B** | C | **B** | A |

Fig. 6. An example to show the benefits of double coding in discovering uncertainty. ICR is equal in both units in existing ICR methods.

*Addressing the Second Challenge: Correlated Uncertainty.* Our goal was to use the valuable details from double coding to help the team quickly and efficiently find the overlapping codes. Thus, we programmed our spreadsheet prototype such that it (1) compared the primary and secondary codes across all coders to identify rows where secondary codes were different than primary codes or rows, (2) determined the correlation between the primary and secondary codes and, (3) displayed the correlations in a color-coded matrix (Fig. 7).

For example, for all the units for which Disparate Systems (purple) was chosen as the primary code (fourth row), the following codes were selected as secondary: Computational Mechanism Issues (pink) 8% of times, Coordination Breakdown (green) 8% of times, and Disparate Systems 85% of times. In other words, 85% of the time, coders were certain about selecting Disparate Systems because they chose the same primary and secondary codes. On the other hand, Coordination Breakdown (green) has no overlap with other codes when assigned as the primary code, which shows that coders were absolutely certain in using this code.

Code Wizard automatically generated this information, which helped the team identify the overlapping codes quickly and efficiently, and focus their efforts on the problematic codes.

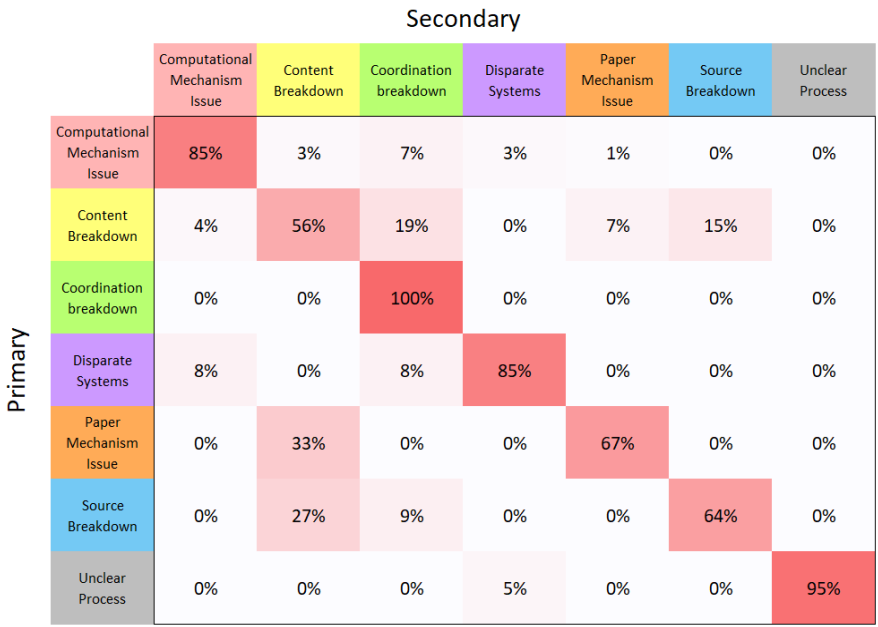

Fig. 7. Correlation between primary and secondary codes

It is important to point out that this matrix (Fig. 7) is not symmetrical. We learned that coders' secondary code assignment depended on the primary code. In other words, when coders selected a well-defined and exclusive code such as Coordination Breakdown (green) as the primary code, they did not experience many difficulties in assigning the same code as the secondary. However, when the primary code was ill defined, such as Content Breakdown (yellow), coders struggled with the secondary code and selected an array of other codes including the well-defined Coordination Breakdown (green).

In addition to generating the aggregated correlations matrix for all coders (Fig. 7), our prototype generated the correlation matrix for each individual coder. The individual coder correlation matrix looks similar to Fig. 7 but uses the data from a single coder. For example, an individual coder could see that she had the most overlap between codes A and C and decide whether she needed to learn more about those codes or bring that up in the team meeting for more discussion. In addition, this coder could compare her coding results with the aggregated results and see how her codes compared with the rest of the team.

This information helped the team to quickly and efficiently find the overlapping codes and address them accordingly. As mentioned earlier, our ICR in the first round of coding was 0.52 so we needed to code a second time to see if taking the above steps had improved our agreement.

*3.3.3 Third Challenge: Persistent Low Agreement Rate.* In the second round, we coded each unit of analysis only one time for two main reasons: (1) we had previously discussed the issues of overlapping and ambiguous codes as well as coders' uncertainty, and tested the revised codes in a quick sample coding, (2) all ICR methods merely consider the primary code in their formula, so double coding would not have made a difference at this step, and (3) at this point, we had not seen the potential impacts or benefits of measuring coders' uncertainty and double coding on increasing ICR. So we decided not to spend extra time on double coding,

|   | Time | Unit of Analysis | Link to Source | John Primary | Ava Primary | Suzie Primary | Robert Primary | Agreement (Pi) |
|---|---|---|---|---|---|---|---|---|
| 1 | 10:08 6/7/16 | It is not clear if the information is from a citizen or someone else. | go to | SB | SB | SB | SB | 1.00 |
| 2 | 10:08 6/7/16 | The agency that the information is coming from is not on the form. | go to | CT | SB | SB | CT | 0.33 |
| 3 | 10:08 6/7/16 | I ask about this and the player shows me other written forms that similarly require a lot of guessing because the way they are filled out is vague. | go to | CT | CT | CT | CT | 1.00 |
| 4 | 10:20 6/7/16 | One of the planning team goes to track down who handed off two confusing paper information forms. Guesses are made based on pen color and handwriting as to who to talk to. | go to | CT | CT | PM | SB | 0.50 |
| 5 | 10:20 6/7/16 | Another confusing message doesn't have a log number from the message center, so it was not routed through the center. So, the planning players are unsure who to go back and get information from. | go to | CT | CT | PM | SB | 0.17 |

Fig. 8. A screenshot of our second round of coding and the agreement kappa per row

After finishing the second round of coding, our prototype merged the code assignments from individual coders and calculated the ICR. Although our ICR had improved from the first round and went up to 0.73 from 0.52, it was still below the acceptable kappa (0.8).

In addition to color-coding, we included the agreement rate per unit of analysis at the end of each row (Fig. 8). This technique helped us quickly locate the disagreements per line and see the degree of disagreement. Based on Fig. 8, some codes still seemed to overlap so we needed to dig deeper to find the extent of code overlaps and the reasons behind the overlaps.

*Addressing the Third Challenge: Correlated Disagreement Matrix.* Building on the literature, we attempted to address this challenge by leveraging the disagreement in code assignments. Barbour (2001) argues that "the content of disagreement" is more important than the degree of concordance between coders [14]. Lasecki et al. (2014) developed Glance, a tool that measures coders' disagreements to detect problematic or ambiguous analysis queries [30]. Aroyo et al. (2015) indicate that disagreement can address ambiguity, and systematic disagreement among individuals is evidence for having multiple perspectives [31]. Zade et al. (2018) propose metrics to sort and filter disagreements into diversity and divergence to discover the source of disagreements [32].

We took a step further into addressing the code overlaps by focusing on how coders had assigned the codes. We did this by measuring the number of times coders had assigned different codes to a unit of analysis. This is more easily visualized through an example. Imagine three out of six coders assigned code A and three others assigned code C to Unit 1 (Fig. 9). Now, compare Unit 1 to Unit 2, where three coders assigned code B and others assigned A, C, and D (Fig. 9). For Unit 1, we can infer that there may be a strong overlap between codes A and C, especially if we see similar coding patterns repeat for other units of analysis. However, it is difficult to infer much meaningful correlations from the Unit 2 codes. We designed our prototype to detect these patterns of disagreement.

The analogy we used here is similar to our analogy for addressing the second challenge, but unlike the second challenge, here we exclusively focused on the primary code as opposed to comparing it with a secondary code. We merely included the mutual and bilateral relationships between codes (for example correlation between codes B and D) because considering three or higher level of relationships creates complication and confusion computationally and intuitively.

|        | Coder 1 | Coder 2 | Coder 3 | Coder 4 | Coder 5 | Coder 6 |
|--------|---------|---------|---------|---------|---------|---------|
| Unit 1 | C       | A       | C       | A       | A       | C       |
| Unit 2 | D       | B       | B       | C       | A       | B       |

Fig. 9. An example of meaningful patterns in coding

Now, let us describe how our prototype works. Let's imagine five coders (coders 1-5) assigned four codes (A, B, C, and D) to five units of analysis (Fig. 10-a). In this example, Fleiss' kappa is 0.3, which is well below the acceptable level and indicates low agreement between the coders. We designed our prototype to look for potential code overlaps, taking the following steps:

(1) The prototype first counts the number of code combinations assigned to each unit of analysis, called "code connection degree," which is determined by counting the minimum number of times a combination of two codes is used in each unit of analysis. For example, in Fig. 10-a, only two codes are used for Unit 1: A and C. A is used one time and C is used four times.

Code Wizard counts the number of times these two codes appeared together, looking for connections between AA, CC, and AC.

As can be seen in Fig. 10-b, the code connection degree of AA is 1, CC is 4, and AC is 1 (i.e., minimum number of times that codes A and C are assigned to Unit 1). The code connection degrees between all other codes are zero because they are not used for Unit 1. Below is the formula to calculate the code connection degree:

$$AC_{Unit\ 1} = Min\ (\#A_{Unit\ 1}, \#C_{Unit\ 1}) = Min\ (1,4) = 1 \tag{1}$$

(2) In the second step, the prototype calculates the code connection degree for all units of analysis and provides each code connection in the last row.

| | Coder 1 | Coder 2 | Coder 3 | Coder 4 | Coder 5 | | AA Min (#A, #A) | BB Min (#B, #B) | CC Min (#C, #C) | AB Min (#A, #B) | AC Min (#A, #C) | BC Min (#B, #C) |
|---|---|---|---|---|---|---|---|---|---|---|---|---|
| Unit 1 | C | A | C | C | C | Unit 1 | 1 | 0 | 4 | 0 | 1 | 0 |
| Unit 2 | A | B | B | B | A | Unit 2 | 2 | 3 | 0 | 2 | 0 | 0 |
| Unit 3 | B | A | C | A | A | Unit 3 | 3 | 1 | 1 | 1 | 1 | 1 |
| Unit 4 | C | C | C | C | C | Unit 4 | 0 | 0 | 5 | 0 | 0 | 0 |
| Unit 5 | B | B | B | A | A | Unit 5 | 2 | 3 | 0 | 2 | 0 | 0 |
| | | | | | | **Sum (S)** | **8** | **7** | **10** | **5** | **2** | **1** |

10-a. Coded units of analysis      10-b. Code connection degrees

Fig. 10. A hypothetical example of how code connection degree works

(3) Our prototype automatically generates a matrix based on the code connection degrees obtained in the previous step. Since the frequencies of code assignments are not equal and we are interested in discovering the connections between codes, which are not dependent on frequency, we normalized the connection scores (using geometric mean of origin code connection degrees), which we refer to as "correlated disagreement" and calculate it using the following formula:

$$r_{XY} = \frac{S_{XY}}{\sqrt{S_{XX} * S_{YY}}} \tag{2}$$

For example, the correlated disagreement coefficient between codes A and C is 0.2.

$$r_{AC} = \frac{S_{AC}}{\sqrt{S_{AA} * S_{CC}}} = \frac{2}{\sqrt{8 * 10}} = 0.2 \tag{3}$$

(4) In the fourth step, the results will be generated and displayed in a matrix, as shown in Fig. 11.

Fig. 11-c is the "Correlated Disagreement Matrix"—this matrix represents correlation between mutual codes that is analogous to the correlation matrix in statistics, and it quantifies correlated disagreements among codes.

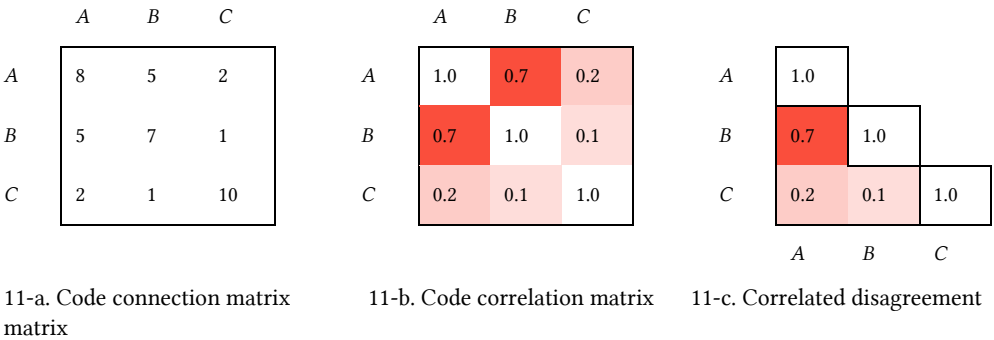

11-a. Code connection matrix  11-b. Code correlation matrix  11-c. Correlated disagreement matrix

Fig. 11. Correlated disagreements between codes of our hypothetical example

Below, we list the main properties of our Correlated Disagreement Matrix:

1. It is symmetrical, so we removed its upper half to reduce visual noise.
2. The value of diagonal cells would always be equal to one. Thus, we removed them from the matrix, as they are not informative.
3. In optimal agreement scenarios (ICR= 1.0), all non-diagonal values would be zero.
4. A high value in non-diagonal cells imply two codes were used interchangeably, which is a strong indication of either an overlap or a misinterpretation of the codes.
5. Higher values of non-diagonal cells (closer to 1) means lower agreement rate, and vice versa.

The example matrix in Fig. 11-c shows high disagreement for codes A and B (0.7). This means the team should investigate these two codes.

As mentioned earlier, Code Wizard automatically performs the calculations and generates the matrix, so coders can focus on other critical aspects of collaborative coding that cannot be automated, such as clarification discussions.

Now that we have explained the mathematics behind generating the Correlated Disagreement Matrix through an example, let's go back to the coding results of our project. As noted earlier, although our team's ICR score improved from 0.52 to 0.73, it was still below the acceptable threshold (0.8). The Correlated Disagreement Matrix generated for our second round of coding (Fig. 12) immediately revealed the problematic areas: coders had the highest correlated disagreement coefficient between Disparate Systems and Computational Mechanism Issue, as well as Content Breakdown and Source Breakdown. Therefore, our team got together to discuss these codes and further revised them.

After team discussions to address ambiguous and overlapping codes and to make our codes exclusive, our team performed a third round of coding to see how our agreement rate changed after taking the above steps.

Our ICR in the third round of coding was 0.81, which is above the acceptable threshold. Our team was satisfied with the agreement rate so we did not hold a meeting to discuss further disagreements. However, Code Wizard generated the Correlated Disagreement Matrix (Fig. 13) to show what changes led to a higher agreement rate.

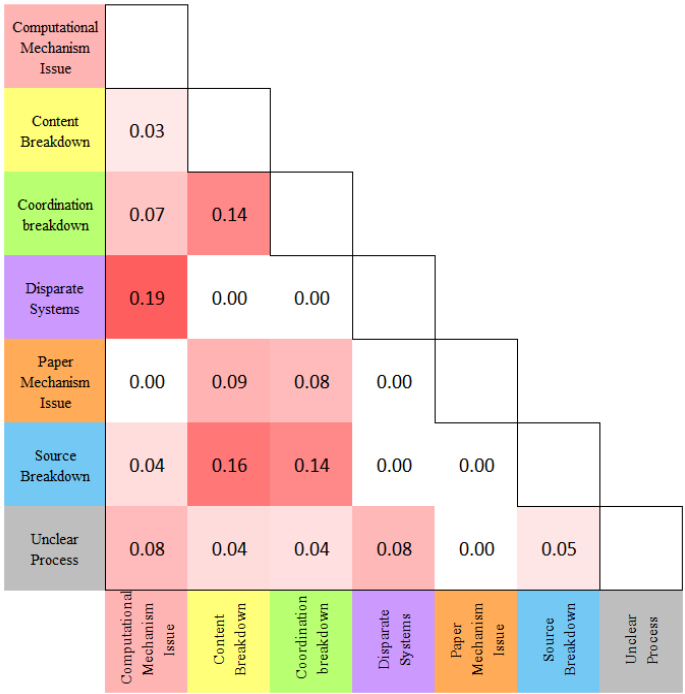

Fig. 12. Correlated Disagreement Matrix generated for second round of coding; Fleiss' kappa = 0.73

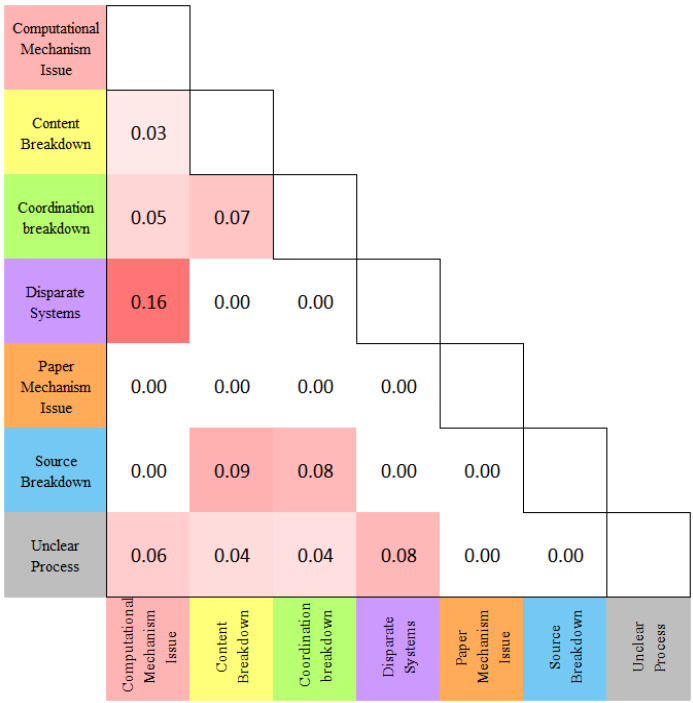

Fig. 13. Correlated Disagreement Matrix generated for third round of coding; Fleiss' kappa = 0.81

As can be seen in Fig. 13, the number of cells with zero value increased, which indicates the exclusivity of those codes. In other words, the problem of high correlation between many codes is addressed here. As Fig. 12 illustrates, in the second round of coding, Paper Mechanism Issues (orange) correlated with both Content Breakdown (yellow) and Coordination Breakdown (green). However, in the third round of coding (Fig. 13), there is no longer any correlation between these codes. In addition, the correlation rate for many codes is reduced, such as for Source Breakdown and Content Breakdown.

In our analysis, we did not add, remove, or alter codes but because Code Wizard is developed in Excel, codes can easily be added, removed, or altered if needed. We programmed Code Wizard to automatically update the codes based on changes in the codebook or where needed.

## 4  DISCUSSION

This study describes the development of a prototype for a coding tool called Code Wizard with the goal of enabling and supporting efficient collaborative coding using visualization techniques and colors.

We discussed three main challenges we encountered throughout the process of collaborative coding that were also expressed by many other scholars, as described in Section 2. We illustrated how Code Wizard helped the process of addressing each challenge and improving our team's collaboration.

Software packages such as NVivo, Dedoose, and Atlas.ti support collaborative coding for qualitative analysis at some levels. In NVivo, Dedoose, and Atlas.ti coding stripes show how different coders coded a piece of data [18, 20, and 21]. In NVivo, the coding comparison query feature enables coders to compare coding of two users or two groups of users and calculates Cohen's kappa and direct agreement percentage between them. Dedoose calculates Cohen's kappa and Pearson's correlation coefficient between two users. However, NVivo and Dedoose do not currently evaluate the ICR coefficient between more than two coders. Atlas.ti does not calculate any type of ICR coefficient. Also, NVivo, Dedoose, and Atlas.ti present the co-occurrence matrix of codes selected for units of analysis simultaneously. NVivo can produce the co-occurrence matrix for codes that assigned by two coders for similar units of analysis too. However, NVivo, Dedoose, and Atlas.ti fall short in creating co-occurrence matrices if the number of coders exceeds two. Code Wizard produces the Correlated Disagreement Matrix, which is conceptually related to the co-occurrence matrix, for any number of coders following the approach presented in Section 3.3.

Code Wizard was gradually developed during the collaborative coding process, so we used the coders' concerns and feedback to improve the tool. For example, the idea of coding a unit of analysis twice came from the coders' discussions in meetings. During the discussion sessions, we realized that coders had difficulties picking one code to assign to the units of analysis. Therefore, we added a feature to the tool to allow them assign two codes that fit best. Having two codes was challenging because ICR coefficients are calculated based on one code. Although this feature would not change the ICR coefficient directly, allowing coders to select two codes and embracing and presenting the uncertainties during the discussion sessions helped the team collaboration.

Code Wizard supported team collaboration in two main areas, the process of coding and the practice of coding. We organized our findings according to these two categories.

**4.1 Process of Coding**

This category includes the mechanisms our teams used to conduct qualitative coding, select a coding tool, perform the coding, measure ICR, and similar activities.

*4.1.1 Collaboration Evolved Over Time.* Code Wizard engaged the team in every step of the process of coding, discussing, and revising the codebook. As the team continually shaped the codebook, our team's collaboration formed around our understanding of the phenomena, which led to further shaping the codebook to create exclusive codes for our research. As Weston et al. indicate (2001), "team builds codes and coding builds a team through the creation of shared interpretation and understanding of the phenomenon being studied" [33]. Code Wizard helped a very diverse team with little experience in the area of disaster management learn and collaborate throughout the coding process in a meaningful way because they could see the impact of their work on the entire process as well as on the research results.

We acknowledge that as the coding team becomes more familiar with the codes and more accomplished at understanding the data and applying the codes, they become faster and better at coding, and potentially their level of engagement can increase. But we should recognize that the coding team changed every 10-12 weeks. While we did not completely control for the new coders, we realized that Code Wizard made onboarding and engaging the new coders quick and easy and the new coders were as engaged as the people who had been on board from the beginning.

*4.1.2 Code Wizard Made the Coding Process Visible.* Collaborative coding includes multiple steps and tasks that are often not talked about or are easy to overlook in qualitative studies. Transparency of the qualitative research process including how the decisions were made and what factors were involved in reaching an agreement are often not discussed in qualitative papers and tend to get buried under reporting the final results. Yet, each of these steps plays a crucial role in ensuring the validity of research findings and ensuring that the codes are not superficial. As Ryan indicates, "it is important to make visible the researcher's decisions and processes. The intricate processes of organising, coding and analysing the data are often rendered invisible in the presentation of the research findings, which requires a 'leap of faith' for the reader." Ryan adds that computer tools can help "make the research process more transparent, without sacrificing rich, interpretive analysis by the researcher" [34]. We programmed Code Wizard to automatically record key decisions, reflections, variations, and emergent ideas for each round of coding in a new spreadsheet. This was possible by comparing visualizations generated from collaborative coding in different rounds. Team members could also see where other coders were in the process, each individual coder's struggles, and how their uncertainties could have a significant impact in the coding process and discussions.

*4.1.3. Coders' Commitment to Data as Opposed to Data Alienation.* Our team's experience diverged from the criticism that using computer-assisted coding programs would alienate researchers from the data [35]. Code Wizard helped the team engage with data in three ways: (1) it made it easy for coders to see in what ways even small changes to the codebook can make a significant impact on the result, (2) it made the coding process transparent so the coders could see where they needed to focus, as opposed to having lengthy and inefficient meetings, which is what we initially experienced with using Dedoose, and (3) Code Wizard provided real-time results of the changes in coding, so coders could interact with data in a more dynamic way.

*4.1.4 Coders' Sense of Ownership of the Results.* Code Wizard generates various visualizations for individual coders that illustrate the impact of their coding on team agreements. Based on coders' feedback during discussion meetings, these informative visualizations increased the individual coders' sense of authority and responsibility, which led to more engagement in the process. In team discussions, individual coders would confidently share new ideas, offer feedback, and talk about improving the results.

*4.1.5 Eliminating the Need for Backtracking the Codes Line by Line.* Code Wizard helped the process of collaborative coding by eliminating the need for going over the units of analyses line by line to find the disagreements. Code Wizard identified and located the primary disagreements so the discussion would be focused on those specific disagreements. Backtracking to locate disagreements can take months to finish. In addition, by the time coders reviewed a few disagreements, they could forget the details of earlier conversations and the earlier reasoning. At times, the clarification discussions would lead to changing the codes or revising the definitions and that would make earlier conversations inapplicable.

## 4.2 Practice of Coding

This category includes how the team communicated and exchanged information at every step in the process.

*4.2.1 Using Uncertainties to Make Communications More Explicit.* We identified certainty/uncertainty as an important factor in the practice of collaborative coding. We measured certainty by asking coders to code each unit of analysis two times in the first coding round. Measuring uncertainty helped team members to not only find ambiguous codes and code overlaps, but also to find the extent to which those codes overlapped. This information helped the team exchange their understanding of the codes and context, and develop actionable plans accordingly.

*4.2.2 Accelerating Coding through Use of Colors and Visualizations.* Using colors proved to be invaluable in locating disagreements and it made discussions quicker and more efficient. We used colors in two main ways: (1) we used distinct colors for codes, which helped in locating the disagreements quickly, and (2) we used shades of red to show the degree of disagreements. Dark red represented significant disagreement, and light pink was used for trivial disagreement. While using colors was desirable by coders and accelerated the coding process, it also created some issues for the team. Despite our best effort to assign distinctive and high-contrast colors to different codes, we realized that colors would appear differently on different computers or operating systems. This unforeseen shift in colors created confusion because at times coders would see different codes being represented by relatively similar colors. In addition, coders initially had difficulties understanding the correlation matrices and visualizations (Figs. 7, 11, and 12). We partially resolved this issue by refining and simplifying the visualizations. We are also experimenting with other types of visualizations in future studies.

*4.2.3 Interacting with Real-Time Results.* Code Wizard automatically and simultaneously generated all the visualizations and applied colors. This feature tremendously helped team discussions because it enabled the team to quickly and easily make changes and see the results in real-time during discussion sessions. The coders could individually or in a group modify their codes in the master Excel sheet and instantly see the results and updated visualizations. Unlike tools such as NVivo, in Code Wizard we did not need to upload a new file if people changed their codes during discussion sessions. One of the experienced researchers on the team wrote, "[Code Wizard] facilitated the reconciliation process by allowing all coders to simultaneously see all experts and each coder's input, making it easy to discuss discrepancies."

*4.2.4 Code Wizard Promoted Diverse Team Integration.* In our project, we worked with a diverse group of people with no experience in emergency management who came from different disciplines (civil engineering, human-centered design & engineering, electrical engineering, and cognitive psychology). Working with a diverse team introduced some challenges in reaching a common language that was understandable by all. Thus, we paid special attention to creating an inclusive and accessible tool that respects differences among a diverse group of coders and supports their comfort level.

*4.2.5 Little Training Required.* Programming Code Wizard's prototype within Excel, a familiar software, meant not much training was required. Team members sent the individual coding to the team lead, who incorporated them into Code Wizard. The rest of the processes are all automated. During discussion sessions, coders can modify their coding in Code Wizard and the result will change automatically in real-time.

As noted earlier, we had initially experienced many difficulties with existing commercial coding tools. We spent more time learning their features than working on our research. Our team expressed feeling relief when Code Wizard was being developed during the course of the research. One of our team members programmed Excel as a platform and this decision was highly supported by the team members because (1) everyone was familiar with Excel, (b) everyone had access to Excel, (c) Excel was free for students—if they did not already have it installed it on their computers, it was available through the university IT center.

## 5 LIMITATIONS

In this work, we focused on improving collaborative coding experience in qualitative research. Our case study showed promising results when multiple coders collaborate on a large dataset made of textual field notes in an inductive qualitative research. Further research will show whether Code Wizard would support other types of qualitative data such as audio, tweets, or other forms of text media. We also do not know the possible benefits or impacts of using Code Wizard on larger teams or teams of two coders.

Code Wizard may not be as helpful when a gold standard coding scheme is already established and when the codes are mutually exclusive, but it may help train the coders. In addition, Code Wizard currently does not resolve the issue of measuring ICR for multiple coding and hierarchical coding, but we look forward to extending Code Wizard to handle multiple coding and hierarchical coding.

In our example, we used Fleiss' kappa, which does not work when no code is assigned to a unit of analysis, but we would like to test Code Wizard for such instances, as either primary or secondary. Finally, we observed that the coders sometimes struggled to understand the visualizations presented in Figs. 7, 11, and 12. We revised these visualizations to make them as simple as possible. We did not test Code Wizard with other types of visualizations such as chord diagram, but we plan to do in future studies. As we improve Code Wizard, we would like to explore ways of extending it to support other types of coding and incorporate more effective visualization techniques.

## 6 CONCLUSIONS

This research is a contribution to an ongoing effort to address the challenges of collaborative coding of qualitative data using computerized tools. We believe that there is a need for a simpler tool to help teams that do not require the heavy weight and sophisticated features offered by commercial tools. Completely unaided collaborative coding is difficult, while commercial tools can add overhead. We should make the distinction between the support offered by a simple and lightweight tool such as Code Wizard versus the support that the heavy weight commercial tools offer. Our goal was to design a middle-level tool for teams with simple coding needs and not to replace the existing tools. We believe there is space to design tools for this middle-level coding, and we hope our effort can open a new avenue for the CSCW community interested in developing simple tools to support collaborative coding.

As qualitative data analysis becomes increasingly important in various fields and with advances in technology, we now need to process data in larger amounts and with more complexity. We believe that visual analytics and computerized coding tools can support the tedious process of collaborative coding more efficiently. Our work simplified the process of detecting ambiguous and overlapping codes and offered insights into the process of measuring ICR. If we combine all the solutions described in this paper, Code Wizard would be able to more methodically manage the collaboration.

## ACKNOWLEDGMENTS

The authors would like to thank Drs. Mark Haselkorn, Mark Zachry, and Sonia Savelli for their invaluable guidance throughout the development of Code Wizard. Also, many thanks to all researchers and coders who participated in our coding collaborations.

## REFERENCES

[1] Klaus Krippendorff. 2011. Agreement and Information in the Reliability of Coding. *Communication Methods & Measures* 5 (2): 93–112.
[2] Angela Sweeney, Kathryn E Greenwood, Sally Williams, Til Wykes, and Diana S Rose. 2013. Hearing the Voices of Service User Researchers in Collaborative Qualitative Data Analysis: The Case for Multiple Coding. *Health Expectations* 16 (4): e89–99. https://doi.org/10.1111/j.1369-7625.2012.00810.x.
[3] Laila Burla, Birte Knierim, Jurgen Barth, Katharina Liewald, Margreet Duetz, and Thomas Abel. 2008. From Text to Codings: Intercoder Reliability Assessment in Qualitative Content Analysis. *Nursing Research* 57 (2): 113. https://doi.org/10.1097/01.NNR.0000313482.33917.7d.
[4] M. Schreier. 2012. Qualitative content analysis in practice. Thousand Oaks, CA: Sage.
[5] Satu Elo, Maria Kääriäinen, Outi Kanste, Tarja Pölkki, Kati Utriainen, and Helvi Kyngäs. 2014. Qualitative Content Analysis: A Focus on Trustworthiness. *SAGE Open* 4 (1): 2158244014522633. https://doi.org/10.1177/2158244014522633.
[6] Catherine Pope, Sue Ziebland, and Nicholas Mays. 2000. Analysing Qualitative Data. *BMJ: British Medical Journal* 320 (7227): 114–16.
[7] Klaus Krippendorff. 2004. Reliability in Content Analysis. *Human Communication Research* 30 (3): 411–33. https://doi.org/10.1111/j.1468-2958.2004.tb00738.x.
[8] Catherine MacPhail, Nomhle Khoza, Laurie Abler, and Meghna Ranganathan. 2016. Process Guidelines for Establishing Intercoder Reliability in Qualitative Studies. *Qualitative Research* 16 (2): 198–212. https://doi.org/10.1177/1468794115577012.
[9] Daniel J. Hruschka, Deborah Schwartz, Daphne Cobb St.John, Erin Picone-Decaro, Richard A. Jenkins, and James W. Carey. 2004. Reliability in Coding Open-Ended Data: Lessons Learned from HIV Behavioral Research. *Field Methods* 16 (3): 307–31. https://doi.org/10.1177/1525822X04266540.
[10] Jan Schilling. 2006. On the Pragmatics of Qualitative Assessment: Designing the Process for Content Analysis. *European Journal of Psychological Assessment* 22 (1): 28–37. https://doi.org/10.1027/1015-5759.22.1.28.
[11] Matthew Lombard, Jennifer Snyder-Duch, and Cheryl Campanella Bracken. 2002. Content Analysis in Mass Communication Assessment and Reporting of Intercoder Reliability. *Human Communication Research* 28 (4): 587–604. https://doi.org/10.1111/j.1468-2958.2002.tb00826.x.
[12] Lynda Berends, and Jennifer Johnston. 2005. Using Multiple Coders to Enhance Qualitative Analysis: The Case of Interviews with Consumers of Drug Treatment. *Addiction Research & Theory* 13 (4): 373–81.

https://doi.org/10.1080/16066350500102237.
[13] Peter Nielsen. 2012. Collaborative Coding of Qualitative Data (White Paper), www.researchgate.net/publication/261633642_Collaborative_Coding_of_Qualitative_Data_White_paper, Accessed April 19th, 2018.
[14] Rosaline S. Barbour. 2001. Checklists for Improving Rigour in Qualitative Research: A Case of the Tail Wagging the Dog? *BMJ: British Medical Journal* 322 (7294): 1115–17.
[15] Ron Artstein, and Massimo Poesio. 2008. Inter-Coder Agreement for Computational Linguistics. *Computational Linguistics* 34 (4): 555–96. https://doi.org/10.1162/coli.07-034-R2.
[16] J. Richard Landis, and Gary G. Koch. 1977. The Measurement of Observer Agreement for Categorical Data. *Biometrics* 33 (1): 159–74. https://doi.org/10.2307/2529310.
[17] https://cat.texifter.com/ Accessed April 19th, 2018
[18] https://atlasti.com/ Accessed April 19th, 2018
[19] Caroline E Wood, Michelle Richardson, Marie Johnston, Charles Abraham, Jill Francis, Wendy Hardeman, Susan Michie, DPhil,, "Applying the behaviour change technique (BCT) taxonomy v1: a study of coder training," Transl Behav Med, vol. 5, no. 2, pp. 134–148, Jun. 2015.
[20] http://www.qsrinternational.com/nvivo/nvivo-products, Accessed April 19th, 2018
[21] http://www.dedoose.com/, Accessed April 19th, 2018
[22] T. Blascheck, F. Beck, S. Baltes, T. Ertl, and D. Weiskopf. 2017. Visual analysis and coding of data-rich user behavior. *IEEE TVCG*, 2017.
[23] https://www.quirkos.com/index.html, Accessed April 19th, 2018
[24] Margaret Drouhard, Nan-Chen Chen, Jina Suh, Rafal Kocielnik, Vanessa Pena-Araya, Keting Cen, Xiangyi Zheng, and Cecilia R. Aragon. 2017. Aeonium: Visual analytics to support collaborative qualitative coding. *In Pacific Visualization Symposium (PacificVis), 2017 IEEE*, pp. 220-229. IEEE, 2017.
[25] https://www.fema.gov/media-library-data/1484078710188-2e6b753f3f9c6037dd22922cde32e3dd/CR16_AAR_508.pdf, Accessed April 19th, 2018
[26] https://www.fema.gov/media-library-data/1462203815175-6b989e683e8c2d34864f007fbde2c3fd/CR2016Flyer.pdf, Accessed April 19th, 2018
[27] https://www.fema.gov/news-release/2017/06/07/emergency-managers-announce-improvements-after-cascadia-rising-exercise, Accessed April 19th, 2018
[28] https://slack.com/, Accessed April 19th, 2018
[29] Bjørn, P., Ciolfi, L., Ackerman, M., Fitzpatrick, G., & Wulf, V. (2016). Practice-based CSCW Research: ECSCW bridging across the Atlantic. Proceedings of the 19th ACM Conference on Computer Supported Cooperative Work and Social Computing Companion, 26, 210-220.
[30] Walter S. Lasecki, Mitchell Gordon, Danai Koutra, Malte F. Jung, Steven P. Dow, and Jeffrey P. Bigham. 2014. Glance: Rapidly Coding Behavioral Video with the Crowd. In *Proceedings of the 27th Annual ACM Symposium on User Interface Software and Technology*, 551–562. UIST '14. New York, NY, USA: ACM. https://doi.org/10.1145/2642918.2647367.
[31] Lora Aroyo and Chris Welty. 2015. Truth is a lie: Crowd truth and the seven myths of human annotation. *AI Magazine* 36, 1 (2015), 15–24.
[32] Himanshu Zade, Margaret Drouhard, Bonnie Chinh, Lu Gan, and Cecilia Aragon. n.d. 2018. Conceptualizing Disagreement in Qualitative Coding, 11. *CHI 2018*.
[33] C. Weston, T. Gandell, J. Beauchamp, L. McAlpine, C. Wiseman, and C. Beauchamp. 2001. Analyzing Interview Data: The Development and Evolution of a Coding System. *Qualitative Sociology*, vol. 24, no. 3, pp. 381–400, Sep. 2001.
[34] Mary Ryan. 2009. Making visible the coding process: Using qualitative data software in a post-structural study. *Issues in Educational Research*, Vol 19(2), 2009
[35] N. G. Fielding, and R. M. Lee. 2002. New patterns in the adoption and use of qualitative software. *Field Methods*, 14(2), 197-216.